\documentclass[11pt, onecolumn]{IEEEtran}
\usepackage{amsmath}
\usepackage{amsfonts}
\usepackage{amssymb}
\usepackage{amscd}
\usepackage[dvips]{graphicx}
\usepackage{epsfig}
\usepackage{subfigure}
\centerfigcaptionstrue
\def\BibTeX{{\rm B\kern-.05em{\sc i\kern-.025em b}\kern-.08em
    T\kern-.1667em\lower.7ex\hbox{E}\kern-.125emX}}

\usepackage{newlfont}

\long\def\symbolfootnote[#1]#2{\begingroup%
\def\thefootnote{\fnsymbol{footnote}}\footnote[#1]{#2}\endgroup}

\newcommand{\beq}{\begin{equation}}
\newcommand{\eeq}{\end{equation}}
\newcommand{\ben}{\begin{enumerate}}
\newcommand{\een}{\end{enumerate}}
\newcommand{\ber}{\begin{eqnarray}}
\newcommand{\eer}{\end{eqnarray}}

\newcommand{\real}{\mathrm{Real}}
\newcommand{\imag}{\mathrm{Imag}}

\title{Cooperative Sequential Spectrum Sensing Algorithms for OFDM}
\author{ArunKumar Jayaprakasam, Vinod Sharma, Chandra R. Murthy$^{*}$ and Prashant Narayanan\\
Dept. of ECE,  Indian Institute of Science, Bangalore 560 012 \\
  arunkumar.jayaprakash@yahoo.in, vinod@ece.iisc.ernet.in, cmurthy@ece.iisc.ernet.in, prashant2413@gmail.com
         }

\begin{document}

\maketitle

\begin{abstract}
This paper considers the problem of spectrum sensing in cognitive radio networks when the primary user employs Orthogonal Frequency Division Multiplexing (OFDM). We develop cooperative sequential detection algorithms based on energy detectors and the autocorrelation property of cyclic prefix (CP) used in OFDM systems and compare their performances. We show that sequential detection provides much better performance than the traditional fixed sample size (snapshot) based detectors. We also study the effect of model uncertainties such as timing and frequency offset, IQ-imbalance and uncertainty in noise and transmit power on the performance of the detectors. We modify the detectors to mitigate the effects of these impairments. The performance of the proposed algorithms are studied via simulations. It is shown that energy detector performs significantly better than the CP-based detector, except in case of a snapshot detector with noise power uncertainty. Also, unlike for the CP-based detector, most of the above mentioned impairments have no effect on the energy detector.
{\symbolfootnote[0]{This work was supported in part by a grant from the Aerospace and Networking Research Consortium. The initial part of this work on CP detectors has been submitted to the IEEE ICC 2010.}}
\end{abstract}

\newpage
\section{Introduction}
Cognitive Radios, also called secondary users, use the radio spectrum licensed to other (primary) users. They perform radio environment analysis, identify the spectral holes and then operate in those holes \cite{ref1}. Spectrum Sensing (SS), an essential first step in enabling Cognitive Radio (CR) technology, involves (1) identifying spectrum opportunities by detecting holes (white space) when they become available and (2) detecting when the primary reclaims an identified spectral hole. This needs to be done such that the guaranteed maximum interference levels to the primary are not exceeded and there is efficient use of spectrum by the secondary \cite{ref2}. Thus, the cognitive radio needs to detect reliably, quickly and robustly, possibly weak primary user signals. For example, the IEEE 802.22 standard \cite{ref1} requires a sensitivity of -116dBm, while keeping the probability of miss detection under 0.1, using a sensing duration $<$ 2 seconds. 

Orthogonal Frequency Division Multiplexing (OFDM) is used in 802.11a/g wireless LAN's (WLAN), Wireless MAN's (IEEE 802.16 WiMAX,) 3GPP Long Term Evolution (LTE), etc. Because of its widespread acceptance and deployment, it is likely that a primary user would be using OFDM, thus making the problem of detecting OFDM signals especially relevant for cognitive radio. Most of the OFDM systems also employ a cyclic prefix (CP), which implies that the autocorrelation is non-zero at delays of the useful symbol length $-$ a property that can be exploited for spectrum sensing \cite{ref3}. Literature on spectrum sensing is vast, despite it being a relatively recent topic of research. We now briefly summarize the relevant recent work.
\subsection{Literature Survey}
For spectrum sensing, primarily three signal processing techniques \cite{ref4} are proposed in literature: matched filter \cite{ref5}, energy detection \cite{ref5} and cyclo-stationary feature detection \cite{ref6}. Matched filtering is optimal but requires detailed knowledge of the primary signal. When no such knowledge is available, an energy detector is optimal \cite{ref5}. Hence, most of the literature is based on energy detection. However, unlike for the matched-filter and the cyclo-stationary detectors, it suffers from the so-called SNR wall problem in the presence of transmit power or receiver noise power uncertainties (\cite{ref5}, \cite{ref5_b}).

Another important problem encountered in cognitive radios is the hidden node problem caused due to shadowing or time-varying multipath fading. To alleviate this problem, cooperative sequential spectrum sensing algorithms are suggested. Cooperative spectrum sensing where the decisions of different secondaries are fused to obtain the final decision has been studied in \cite{ref4}, \cite{ref7}, \cite{ref8} and \cite{ref9}.  Sequential detection techniques have been used in \cite{ref10}, \cite{ref11}, \cite{ref12} and \cite{ref13}.

Spectrum sensing in an OFDM environment has been studied in \cite{ref3}, \cite{ref14} and for Orthogonal Frequency Division Multiple Access (OFDMA) systems in \cite{ref15}, \cite{ref16}. In \cite{ref14} CP correlation based snapshot and sequential detectors are studied. In \cite{ref15} the effect of time and frequency offset on CP based snapshot detector is studied. In \cite{ref16}, the joint sensing and channel scheduling problem is addressed for OFDM-based CR systems.
\subsection{Our Contribution}
In this paper, we provide spectrum sensing algorithms (for detecting spectral holes in time) when the primary is using OFDM. As in recent work on spectrum sensing in OFDM, we exploit the autocorrelation property in our spectrum sensing algorithms. We compare this with energy detector based algorithms. Furthermore, we study how some of the common impairments in a secondary receiver \cite{ref17} like frequency offset, timing offset, IQ-imbalance \cite{ref18} and uncertainty in the noise power and unknown channel gains affect the statistics, and thus the performance of these detectors. We propose techniques to modify the detectors to mitigate some of the losses. For simplicity, these techniques are first presented in the context of fixed sample-size detectors, but are then extended to the sequential detection setup as well. Here, we are primarily interested in the sequential detection algorithms which are more efficient than fixed sample size (snapshot) detectors, commonly used in the literature. In particular, we use DualCUSUM, a distributed, cooperative sequential-detection algorithm developed in \cite{ref19}. It was shown in \cite{ref13} that DualCUSUM outperforms several other existing spectrum sensing algorithms. However in \cite{ref13} it was not studied in an OFDM context and only the energy detector was studied.

We show that, unlike for the CP-based detector, the energy detector is inherently robust to timing offset, frequency offset and IQ imbalance. However, it is sensitive to noise power uncertainties, while the CP based detector is not. In both the cases, we modify the detector and improve the performance in the presence of impairments. Our study shows that energy detector substantially outperforms CP-based detectors in the setup considered here (except in case of snapshot detector with noise power uncertainties).

This paper is organized as follows. Section II describes the OFDM model. We also present our cooperative sequential detection based setup. In Section III we study the effect of the impairments on the snapshot CP-detector and present possible techniques to overcome the same. In Section IV we consider the energy detector in a snapshot setup and study the effect of the impairments. In Section V we extend these techniques to the sequential change detection algorithms. Section VI concludes the paper.
\section{Model}
We consider a CR network which is sensing a primary using an OFDM system. The OFDM system of the primary (Figure 1) consists of $L_d$ narrowband signals $D_0,D_1,...,D_{L_d-1}$ carried by the subcarriers. An OFDM symbol is obtained by passing the $L_d$ signals through an inverse fast Fourier transform (IFFT). In addition, a cyclic prefix of length $L_c$ is also appended to make the total OFDM symbol duration $L_s=L_d+L_c$. The inter-carrier spacing is denoted by $\triangle f$.

  \begin{figure}[ht]
  \begin{center}
  \epsfxsize=6.5in
  \includegraphics[scale=0.75]{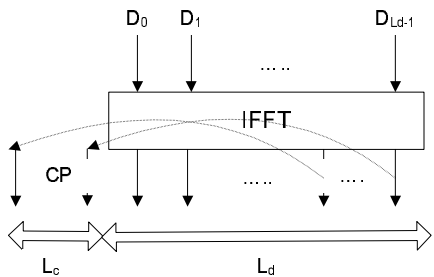}
  \caption{OFDM symbol construction.}
  \label{fig1}
  \end{center}  
  \end{figure}

Define, $d(k)=\sum_{n=0}^{L_d-1}D_ne^{j2\pi k\frac{n}{L_d}}$. The baseband OFDM signal at time $k$ is, 
\beq \label{eq1}
S(k) =  \left\{ \begin{array}{cc}
d(k+L_d-L_c), & k=0,1,...,L_c-1,\\
d(k-L_c), & ~~~~~~~~k=L_c,L_c+1,...,L_s-1.\\
\end{array} \right.
\eeq
By the Central Limit Theorem (CLT), $S(k)$ is approximately Gaussian, since it is a linear combination of $L_d$ random signals. Also, as $E[D_n]=0$, $E[S(k)]=0$. We will take $L_d=64$, $L_c=16$ and $\triangle f=\frac{10}{64}MHz$ in our simulations, although our analysis and algorithms are general. These parameters are assumed to be known to the secondary systems.

In (\ref{eq1}), we observe that the symbols $S(0),...,S(L_c-1)$ are repeated as $S(L_s-L_c),...,S(L_s-1)$. Thus, if we correlate this sequence with a shift of $L_d$, we will get a good correlation in case the primary is transmitting; otherwise not. CP based detectors exploit this property in detecting the primary signal \cite{ref3}.

We consider a cognitive radio system with $L$ secondary users that sense a channel via CP-detectors. Later on, we will also consider energy detectors. The observations made on the channel by these secondaries are processed and sent to a fusion center, which makes a decision on whether the channel is free or not. Then, that decision is sent to all the secondary users for possible use of the channel. The secondary system has to detect when the primary starts transmission (OFF$\rightarrow$ON) and when it stops (ON$\rightarrow$OFF). In the following, we explain our setup for OFF$\rightarrow$ON, but our algorithms work for ON$\rightarrow$OFF also. 

Let the primary start transmission at a random time $T$. Then at time $k$ the signal received by the $l^{th}$ secondary is,
\beq \label{eq2}
X(k,l) =  \left\{ \begin{array}{cc}
N(k,l), & ~~~~~~k=1,2,\ldots,T-1,\\
h(l)S(k)+N(k,l), & ~~~k=T,T+1,\ldots.\\
\end{array} \right.
\eeq

\noindent where $h(l)$ is the channel gain of the $l^{th}$ user,  and $N(k,l)$ is observation noise at the $l^{th}$ user. We assume that the fading is frequency flat and remains constant during the interval of observation (say, approximately, for a duration of ON/OFF period). Slow fading scenarios with primary staying ON and OFF for a few seconds will approximately satisfy this. This assumption is commonly made in the literature \cite{ref8}, \cite{ref11}. (We will see below that most of our algorithms can be used for the frequency selective fading as well). We also assume that $\{d(k), k\geq1\}$ and $\{N(k), k\geq1\}$ are independent and identically distributed (i.i.d.) sequences independent of each other and $T$. Thus, pre-change $X(k,l)$ are i.i.d. with distribution $N_c(0,\sigma^2_{w,l})$, where $N_c$ denotes circularly symmetric complex-normal distribution and $\sigma^2_{w,l}$ the noise power at node $l$ ( $N$ denotes normal distribution). The post-change distribution of $X(k,l)$ is $N_c(0,\sigma^2_{w,l}+\sigma^2_{s,l})$ where $\sigma^2_{s,l}$ is the received power of primary node at node $l$. The effect of different channel gains is absorbed into $\sigma^2_{s,l}$. 

The aim is to detect the change (at random time $T$) at the fusion center as soon as possible at a time $\tau$ ($\geq T$) (i.e., to minimize $E[(\tau-T)^+]$, where $(x)^+ = \max(0,x)$) using the messages transmitted from the $L$ secondaries, with an upper bound on probability of false alarm, $P_{FA} \triangleq P(\tau<T)\leq\alpha$. For this, each of the $L$ nodes uses its observation $X(k,l)$ to generate a signal $Y(k,l)$ and transmits to the fusion center. The data received at the fusion center is corrupted by the additive white Gaussian noise (AWGN) at the receiver. The fusion center uses the observations $Y(k,1),~ ...~,Y(k,L)$ to decide between the two hypotheses $H_0$ (primary not transmitting) and $H_1$. If $H_0$ is chosen, the secondaries continue to use the channel in slot $k$ and the spectrum sensing session continues (they may use part of each slot for sensing and the rest for transmission). If $H_1$ is detected, the secondaries typically switch over to an alternate channel. Our algorithms do not change (although the parameters can) if we interchange the role of $H_0$ and $H_1$. To transmit $Y(k,1),...,Y(k,L)$ from the $L$ secondaries to their fusion node, they need a Multiple Access Control (MAC) protocol. Time Division Multiple Access (TDMA) is the most commonly used protocol.

We have developed (see \cite{ref10}, \cite{ref19}) a robust cooperative algorithm for spectrum sensing in this setup. In this paper, we study this algorithm in the OFDM setup. We also modify it to take care of the various impairments commonly encountered in OFDM.

The rest of the paper is organized as follows. First, in Section III, we study the CP-detector under the classical snapshot setup in the presence of different impairments. In Section IV we study the energy detector under different impairments. In Section V we adopt the different estimation schemes used in Section III and Section IV, to the cooperative sequential detection setup of DualCUSUM to detect the presence of the primary signal. Further, we compare the performance of the energy detector with the CP-detector.

\section{Cyclic Prefix Based Detector}
In this section, we explain the CP correlation based snapshot detector in the context of a single secondary node (thus the subscript $l$ will be omitted in the notation) and present how we mitigate the effects of different impairments and uncertainties. Given a number of observations $X(1),~...~,X(ML_s)$ from $M$ slots of OFDM symbols, we want to detect if $H_0$ or $H_1$ is true. We use the Neymon-Pearson (NP) method for detection. We compute the autocorrelation at lag $L_d$,
\begin{equation}
R = \frac{1}{ML_c} \sum_{j = 0}^{M-1} \sum_{i = 1}^{L_c} X(jL_s+i)X^*(jL_s+i+L_d)
\label{eq3}
\end{equation}
where $X^*$ is the complex conjugate of $X$. The above detector assumes perfect OFDM symbol level synchronization and thus correlates only the exact set of samples which would be repeated in the CP under $H_1$. Using the CLT, it can be shown \cite{ref3} that $R \sim N_c(0,\sigma_0^2)$ under $H_0$ and $R \sim N_c(\sigma_s^2,\sigma_1^2)$ under $H_1$, where $\sigma_0^2 = \frac{\sigma_w^4}{ML_c}$ and $\sigma_1^2 = \frac{(\sigma_w^2+\sigma_s^2)^2}{ML_c} $. At low SNR (i.e., $\sigma_s^2 \ll \sigma_w^2$), $\sigma_0^2 \approx \sigma_1^2$. We work under this assumption, as in CR our main concern is signal detection at low SNR.\\
\indent Since the post-change mean is real, under low SNR conditions, detection is based on the real part $R_r = \real(R)$ as $R_r \sim N(0,\sigma^2)$ under $H_0$ and  $R_r \sim N(\sigma_s^2,\sigma^2)$  under $H_1$, where $\sigma^2 \approx \dfrac{\sigma_w^4}{2ML_c}$. The detection rule is of the form $R_r > \lambda$ for declaration of $H_1$.

In case of frequency selective fading the distribution of $R$ under $H_0$ remains as above. Under $H_1$, $R$ is still approximately Gaussian with mean and variance now depending on the fading parameters. Knowing the fading parameters, one can obtain the detection rule via NP lemma.

Next, we discuss different impairments and possible techniques to mitigate their effects. 

\subsection{Timing Offset}
Timing offset occurs because the cognitive receiver may not know where the OFDM symbol boundary starts in the received set of samples. Thus, it may not know the exact set of samples to correlate in (\ref{eq3}). If it correlates at an incorrect position, $E[R] \approx 0$ under $H_1$. One possible way to take care of this is to correlate for the duration of the entire OFDM symbol:
\begin{equation}
R_r = Re \left( \frac{1}{ML_s} \sum_{i = 1}^{ML_s} X(i)X^*(i+L_d)\right).
\label{eq4}
\end{equation}
Now $\sigma^2 \approx \dfrac{\sigma_w^2}{2ML_s}$ under either hypothesis. The post change mean under $H_1$ is $\mu \sigma_s^2$, where $\mu = \dfrac{L_c}{L_s}$. Because of this, one can expect the performance to degrade. Using $M = 100$ and $SNR = -10dB$ ($\sigma_w^2 = 20, \sigma_s^2 = 2$), we simulated this setup to show the effects of timing offset. We use these parameters throughout this section. The unknown timing offset is chosen as $30$ samples. For different values of the probability of false alarm $p_{fa}$ (detecting $H_1$ while $H_0$ is true), the detection probability $p_{d}$ (detecting $H_1$ while $H_1$ is true) is shown in Table \ref{table1}. To regain some of the lost performance, instead of correlating over the entire set of samples, we can estimate the unknown timing offset $\theta$ by a maximum likelihood estimator (MLE) \cite{ref17} as:
\begin{equation}
\hat{\theta}_{ML} = \arg\max_{\theta \in \{1,2,\ldots,L_s-1 \}} \{ Re(R(\theta)) - \omega P(\theta) \},
\label{eq5}
\end{equation}
where
\begin{equation*}
R(\theta) = \sum_{j = 0}^{M-1} \sum_{i = 1}^{L_c} X(jL_s+i+\theta)X^*(jL_s+i+L_d+\theta),
\end{equation*}
\begin{equation*}
P(\theta) = \frac{1}{2}\sum_{j = 0}^{M-1} \sum_{i = 1}^{L_c} (|X(jL_s+i+\theta)|^2+ |X(jL_s+i+L_d+\theta)|^2), \text{and} \quad \omega = \frac{\sigma_s^2}{\sigma_s^2+\sigma_w^2}.
\end{equation*}

\indent Under low SNR, $\omega$ is small. Also, for a large number of OFDM symbols, $P(\theta) \approx ML_c(\sigma_s^2+\sigma_w^2)$ under $H_1$ and $P(\theta) \approx ML_c \sigma_w^2$ under $H_0$. Thus, $\omega P(\theta)$ does not affect the max operation in (\ref{eq5}), and we use the simplified estimator
\begin{equation}
\hat{\theta}_{ML} = \arg\max_{\theta \in \{1,2,\ldots,L_s-1 \}} \{ \real(R(\theta)) \}.
\end{equation}

\begin{figure}[htp]
\begin{center}
\includegraphics[scale=0.35]{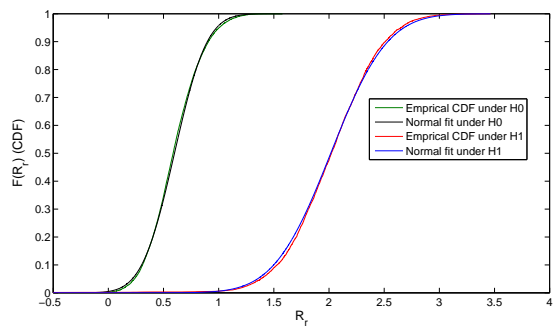}
\caption{Empirical CDFs of $R_r$ under $H_0$ (signal absent) and $H_1$ (signal present) with timing offset. Note that the Gaussian approximation provides a good fit to the empirical CDF.}
\label{Fig2}
\end{center}
\end{figure}

\noindent This estimator has the advantage of not requiring knowledge of $\sigma_s^2$ or $\sigma_w^2$. Then we use the decision statistic
\begin{equation}
R_r = \real \left( \frac{1}{ML} \sum_{j = 0}^{M-1} \sum_{i = 1}^{L_c} X(jL_S+i+\hat{\theta}_{ML}) X^*(jL_S+i+L_d+\hat{\theta}_{ML}) \right)
\label{eq7}
\end{equation}
instead of (\ref{eq3}). Under $H_0$ and $H_1$, now $R_r$ is no longer normally distributed, but the Gaussian distribution still provides a good fit: the empirical distribution of $R_r$ under either hypothesis and the normal fit is shown in Fig.\ref{Fig2}. Thus, we use this approximation for designing the detection threshold and performance analysis. However, as the variances under $H_0$ and $H_1$ are different, the optimal likelihood ratio is not a linear function of $R_r$ and involves knowledge of $\sigma_s^2$ at the CR, which is not desirable. Thus, we propose to continue to use a test of the form $R_r > \lambda$ which is sub-optimal in this case, and could be viewed as a non-parametric test. The performance comparison is shown in Table \ref{table1}. It can be seen that we recover most of the performance lost due to timing offset.

\vspace{-1.1cm}
 
\begin{table}
\begin{center}
\begin{tabular}[t]{|c|c|c|c|}\hline
  $p_{fa}$ & \small{No Impairments} & \small{Correlating over } & \small{Timing Offset} \\
   & \small{(\ref{eq3})} & \small{entire symbol (\ref{eq4}}) & \small{estimate (\ref{eq7})} \\ \hline

  $0.05$ & $0.9999$  &$0.7921$  &$0.9975$  \\ \hline
  $0.025$ & $0.9996$  &$0.7010$ &$0.9944$   \\\hline
  $0.01$ & $0.9988$   &$0.5770$  &$0.9880$  \\\hline
\end{tabular}\caption{snapshot CP detector: Effect of timing offset in $p_d$.The unknown offset is set as 30.}
\label{table1}
\end{center}
\end{table}

\subsection{Frequency Offset}
\indent Let us now consider the scenario when only a frequency offset is present (i.e., the timing offset is assumed to be known). Let the frequency offset (between the cognitive receiver oscillator and the primary transmitter oscillator) be denoted by $\phi$, normalized with respect to the carrier spacing $\Delta f$ . The received signal can be written as $X(k) = S(k) e^{(\frac{j2 \pi \phi k}{L_d})}+N(k)$. Under $H_1$, $R \sim N_c(\sigma_s^2 e^{-j2 \pi \phi}, \sigma_1^2)$. If the receiver is not aware of the frequency offset, the post change $R_r \sim N_c(\sigma_s^2 cos(2 \pi \phi), \sigma^2)$, degrading the performance (see Table \ref{table2}, for $\phi = 0.1$). To mitigate this effect, we estimate the frequency offset $\phi$ via an MLE $\hat{\phi}_{ML}$. The log likelihood ratio can be shown to be proportional to 
\begin{equation}
2 \sigma_s^2(R_r \cos(2 \pi \phi)+ R_i \sin(2 \pi \phi))/\sigma_1^2
\end{equation}
where $R_r$ and $R_i$ are the real and imaginary parts of $R$, respectively. It can be shown that $\hat{\phi}_{ML}=-\angle R/2 \pi $, and we use this estimate in the NP test. Thus, the optimal test becomes $|R|^2 > \lambda '$.  Under $H_0$, $|R|^2$ has an exponential distribution, and under $H_1$, it has a non-central Chi-square distribution. The performance is shown in Table \ref{table2}. Note that once again, most of the performance loss is recovered.

When both timing and frequency offset are present, one can estimate these as
\begin{equation}
\hat{\theta}_{ML} = \arg \max_{\theta} {|R(\theta)|}, \qquad \hat{\phi}_{ML}= - \frac{1}{2 \pi} \angle R(\hat{\theta}_{ML}). \label{eq9}
\end{equation}
We will use these estimates when we consider all impairments together.
\subsection{IQ-Imbalance}
IQ-imbalance occurs due to non-ideal front end components in the receiver \cite{ref18} resulting in the amplitude and phase imbalance in the inphase (I) and quadrature (Q) components of the signal. In the presence of IQ-imbalance the actual received signal is written as
\begin{equation}
 X(k) = \alpha Y(k) + \beta Y^*(k) 
\label{eq10}
\end{equation}
where and $Y(k) = S(k) + N(k), \alpha = \cos(\Delta \phi) +  j\epsilon \sin(\Delta \phi); \beta  = \epsilon\cos(\Delta \phi) - j\sin(\Delta \phi)$ and $\epsilon$ and $\Delta \phi$ are the amplitude and phase imbalance parameters respectively. It can be shown that in the presence of IQ-imbalance,
\beq \label{eq11}
R_r \sim  \left\{ \begin{array}{cc}
N(0, \sigma^2_{IQ}), & ~under H_0,\\
N((1+\epsilon^2)\sigma^2_s, \sigma^2_{IQ}), & under H_1\\
\end{array} \right.
\eeq
where $\sigma^2_{IQ} \approx \sigma^4_w ((1+\epsilon^2)^2 + 4|C_1|^2)/2ML_c$ under low SNR conditions and $C_1 = \alpha \beta^*$. The performance of the detector is shown in Table \ref{table2} for $\Delta \phi = 10^o; \epsilon = 0.2 $. We see that the performance of the detector degrades slightly even when knowledge of imbalance parameters are assumed but not compensated for. However, we can improve performance by compensating for the imbalance. We use the algorithm in \cite{ref18} to compensate for IQ-Imbalance before starting the CP-detector. The imbalance parameters are estimated and corrected for as follows. Let
\begin{equation*}
 \kappa^2 \triangleq \frac{\sum_{i}^{ML_s} X_r^2(i)}{\sum_{i}^{ML_s} X_i^2(i)},  \quad \text{and} \quad \hat\epsilon \triangleq \frac{\kappa-1}{\kappa+1}.
\end{equation*}
where $X_r$ and $X_i$ are the real and imaginary parts of $X$ respectively. Then, one can correct the amplitude imbalance by
\begin{equation}
 Z_r(k) = \frac{X_r(k)}{1+\hat\epsilon}, \qquad Z_i(k) = \frac{X_i(k)}{1-\hat\epsilon}
\label{eq12}
\end{equation}
Assuming the phase imbalance $\in [-\pi/4,\pi/4]$, it is estimated and corrected as,
\begin{equation*}
 \delta = - \frac{\sum_{i}^{ML_s} X_r(i) X_i(i)}{\sum_{i}^{ML_s} (X_r^2(i)+ X_i^2(i))}, \qquad \Delta \hat\phi = \frac{\sin^{-1} (2\delta)}{2}. 
\end{equation*}
Then, instead of using the observations $X(k)$, we use $X'(k)$ with real and imaginary components
\begin{equation}
 {\begin{bmatrix}
X_r'(k) \\
X_i'(k) \\
\end{bmatrix}} = {
\begin{bmatrix}
\cos(\Delta\hat\phi) & \sin(\Delta\hat\phi) \\
\sin(\Delta\hat\phi) & \cos(\Delta\hat\phi) \\
\end{bmatrix}}
{
\begin{bmatrix}
 Z_r(k)\\
 Z_i(k)\\
\end{bmatrix}
}
 \label{eq13}
\end{equation}
for the CP detector. The performance of the detector with this estimator is shown in Table \ref{table2}. We see almost no performance loss.

From these results, we see that the performance loss due to the IQ imbalance could be ignored. However, we have found that it does cause non-negligible degradation when there are other impairments mentioned above. Then the improvement resulting from the compensation procedure described by \eqref{eq12}, \eqref{eq13} can be more significant.

%

\begin{table}
\begin{center}
\begin{tabular}[t]{|c|c|c|c|c|}\hline
  $p_{fa}$ & \small{Frequency Offset} & \small{Frequency Offset} & \small{IQ-Imbalance,} & \small{IQ-Imbalance }
  \\
   & \small{without compensation} & \small{with compensation \eqref{eq9}} & \small{No compensation (\ref{eq11})} & \small{with compensation (\ref{eq12},\ref{eq13})}\\\hline

  $0.05$ & $0.9965$  &$0.9989$
  &$0.9991$  &$0.9999$
  \\\hline

  $0.025$ & $0.9913$  &$0.9975$
  &$0.9977$  &$0.9996$
  \\\hline

  $0.01$ & $0.9794$
  &$0.9939$
  &$0.9937$
  &$0.9988$
  \\\hline
\end{tabular}\caption{Snapshot CP detector: $p_d$ under Frequency Offset and IQ-Imbalance. The normalized frequency offset was set to 0.1 and the IQ imbalance parameters $\Delta \phi$ and $\epsilon$ are set to $10^o$ and 0.2 respectively.}
\label{table2}
\end{center}
\end{table}

\subsection{Noise/Transmit Power Uncertainty}
In a cognitive radio setting, the receiver noise power $\sigma^2_w$ and the received signal power $\sigma^2_s$, may often not be precisely known to the CR \cite{ref5}. We now address the detection problem under these uncertainties. Since the variance of $R_r$ is dependent on the noise power, the detection threshold cannot be set without its knowledge at the CR receiver. Thus, the noise power is estimated as
\begin{equation}
\hat \sigma^2_w = var(X) \approx \frac{\sum_{i=1}^{ML_s} X(i) X^*(i)}{ML_s}
\label{eq14}
\end{equation}
and this is used to set the threshold $\lambda$ to achieve desired $p_{fa}$. This causes a minor performance loss if this estimate is obtained when $H_1$ is true, since, then $var(X) \approx (\sigma^2_w + \sigma^2_s)^2/ML_s$. However at low SNR's this causes small estimation error. This can be verified from Table \ref{table3}. Also, we have been using tests of the form $R_r > \lambda$ or $|R_r|>\lambda$ (partly motivated by the constraints of the present section), and the statistics of $R_r$ do not depend upon $\sigma_s^2$ under $H_0$. Thus, knowledge of receive signal power is not necessary to set the threshold  $\lambda$ to achieve the desired $p_{fa}$.

\subsection{All Impairments}
In this section, we simulate the performance of the fixed sample size CP-detector when all impairments are present. First, the detector estimates and compensates for IQ imbalance using \eqref{eq12} and \eqref{eq13}. Then, the variance of received signal is estimated to set the threshold. Next, the optimal timing and frequency offsets are estimated using \eqref{eq9} and the test is of the form $|R|>\lambda$. The performance is shown in Table \ref{table3}, under the impairments and data statistic given in this section. We see that the estimation schemes recover most of the losses.

For reference, we also compare with the detector in \eqref{eq4} in the presence of IQ-imbalance, frequency offset and noise uncertainty. Noise uncertainty for this detector is taken care of as in Section III.D (i.e., estimating the noise variance to adjust the threshold) as this is necessary to set the threshold. It takes care of timing offset by correlating over the entire OFDM symbol duration, but the detector is unaware of frequency offset and IQ imbalance. Thus, even with partially compensating for the impairments, the performance can be very poor. However, from the last column in Table \ref{table3}, we see that using the methods presented here, most of the losses can be recovered.
Motivated by these estimation schemes, we mitigate the effects of these impairments in the sequential detection algorithm, DualCUSUM, in Section V for the CP-detector. 
\begin{table}
\begin{center}
\begin{tabular}[t]{|c|c|c|c|c|}\hline
  $p_{fa}$ & \small{Noise power} & \small{All impairments} & \small{All impairments} 
  \\
   & \small{estimation \eqref{eq14}} & \small{(\eqref{eq4} with noise power estimation)} & \small{with all compensation} \\\hline

  $0.05$ & $0.9999$  &$0.5122$
  &$0.9712$  
  \\\hline

  $0.02$ & $0.9995$  &$0.3908$
  &$0.9556$  
  \\\hline

  $0.01$ & $0.9976$
  &$0.2614$
  &$0.9300$
  
  \\\hline
\end{tabular}\caption{snapshot CP detector: $p_d$ under noise uncertainty and all impairments}
\label{table3}
\end{center}
\end{table}

\section{Energy Detector}
In this section, we study the performance of the energy detector under a snapshot setup for a single secondary node (as in Section III). We study the effect of different impairments and explore possible techniques to mitigate the same. We compute the energy 
\begin{equation}
 V = \frac{1}{ML_s} \sum_{i=1}^{ML_s} |X(i)|^2.
\end{equation}
Using the CLT, it can be shown that 
\beq \label{eq_V_H0_H1_NoImp}
V \sim \left\{ \begin{array}{cc}
N\left(\sigma_w^2, \frac{\sigma_w^4}{M L_s}\right), & \text{under}~~ H_0,\\
N\left((\sigma_w^2+\sigma_s^2), \frac{(\sigma_w^2+\sigma_s^2)^2}{M L_s} + \frac{2L_c\sigma_s^4}{ML_s^{2}}\right), & \text{under}~~ H_1.\\
\end{array} \right.
\eeq
The additional term in variance of $V$ under $H_1$ arises due to the presence of the cyclic prefix. But at low SNR assumptions, it is easy to see that $V$ is approximately distributed as $\sim N(\sigma_s^2 + \sigma_w^2,\frac{(\sigma_s^2 + \sigma_w^2)^2}{ML_s})$  under $H_1$. We work under this assumption. All the likelihood ratio tests are of the form $V > \lambda$, as the likelihood ratio test will involve the knowledge of primary signal power $\sigma_{s}^2$. 

For the frequency selective case, as in CP detector, $V$ will again be approximately Gaussian with the mean and variance under $H_1$ different from the frequency flat case.


\subsection{Timing Offset} \label{sec:tim_offset}
The effect of timing offset in the context of energy detection is that in a set of $ML_s$ samples, we do not know exactly how many samples would belong to the cyclic prefix portion of the OFDM symbol. This in turn implies that we would not know exactly how many of the terms in the expression for $V$ given by \eqref{eq_V_H0_H1_NoImp} would be correlated. For example, for a
timing offset of $\theta \in \{L_c, ... ,L_s-L_c-1\}$,
\beq \label{eq_en_T_tim_off}
V \sim N \left(\sigma_s^2+\sigma_w^2, \frac{(\sigma_s^2+\sigma_w^2)^2}{M L_s} + \frac{2(M-1)L_c \sigma_s^4}{M^2 L_s^2}\right),
\eeq
i.e., the second term in variance could be different from that given by under $H_1$ could be different from that given by \eqref{eq_en_T_tim_off}, depending upon the value of $\theta$. But under low SNR conditions the effect of this is negligible, and thus timing offset does not affect the performance of the energy detector. The results are shown in Table \ref{table4}. The parameters are $M = 40$ , $SNR = -10$dB
$(\sigma_w^2 = 20,\sigma_s^2 = 2)$ and the unknown timing offset was chosen as 30. The number of OFDM symbols used in this section is different from that in Section III. This is because, with $M=40$ OFDM symbols, the energy detector provides a much superior performance compared to the CP detector, under no noise uncertainty.

\subsection{Frequency Offset} \label{sec:freq_offset}
As the effect of frequency offset is a rotation of $X(k)$ and since the distribution of $X(k)$ is rotationally invariant, the statistics of $V$ is not affected by the frequency offset. Thus, the performance of the Energy detector is not affected by frequency offset. The assumption here is that the loss in signal energy due to the implicit band pass filtering prior to energy detection is negligible. The results are shown in Table \ref{table4} for a frequency offset of $\phi  = 0.1$ (normalized by the inter carrier spacing $\Delta f$).
\begin{table}
\begin{center}
\begin{tabular}[t]{|c|c|c|c|}\hline
  $p_{fa}$ & \small{No Impairments} & \small{Timing offset} & \small{Frequency Offset}
  \\
   & (\ref{eq13}) & & \\\hline

  $0.05$ & $0.9999$  &$0.9999$
  &$0.9999$
  \\\hline

  $0.025$ & $0.9996$  &$0.9997$
  &$0.9996$
  \\\hline

  $0.01$ & $0.9988$
  &$0.9989$
  &$0.9988$
   \\\hline
\end{tabular}\caption{Snapshot Energy Detector: $p_d$ under timing offset=30 and normalized frequency offset=0.1}
\label{table4}
\end{center}
\end{table}

\subsection{IQ-Imbalance} \label{sec:IQ_IMB}
In the presence of IQ-imbalance, the statistics of energy detector are:
\beq \label{eq:T_H0_H1}
V \sim \left\{ \begin{array}{cc}
N\left((1+\epsilon^2)\sigma_w^2, \frac{\sigma_w^4\left((1+\epsilon)^2+4 |\alpha|^2|\beta|^2\right)}{M L_s}\right), & \text{under}~~ H_0,\\
N\left((1+\epsilon^2)(\sigma_w^2+\sigma_s^2), \frac{(\sigma_w^2+\sigma_s^2)^2\left((1+\epsilon)^2+4 |\alpha|^2|\beta|^2\right)}{M L_s}\right), & \text{under}~~ H_1.\\
\end{array} \right.
\eeq
The performance of IQ imbalance under no compensation and with the compensation scheme of Section III.C is shown in Table \ref{table5}.

\subsection{Noise/Transmit Power uncertainty} \label{sec:NS_uns}
It is well known that under presence of noise uncertainty, the energy detector has a $SNR$ wall and the performance suffers. This is illustrated in this subsection. Let $\sigma_w^2 \in [\frac{\bar{\sigma}_w^2}{\delta},\bar{\sigma}_w^2 \delta]$, where $\delta$ denotes the uncertainty level and
$\bar{\sigma}_w^2$ denotes the nominal noise power
used in other sections. The performance of the energy detector when $\bar{\sigma}_w^2 =10$ and $\delta = 1.08$ (corresponding to 0.33 dB) is shown in Table 5. The energy detector sets the threshold for $\bar{\sigma}_w^2 \delta$ and thus the probability of detection significantly degrades. Also, as mentioned in Section III, since the tests are of the form $V > \lambda$, knowledge of $\sigma_s^2$ is not necessary to meet a desired $p_{fa}$.

\subsection{All Impairments}
In this section, we simulate the performance when all impairments excluding noise uncertainty are present (IQ-imbalance is compensated) and then later include the effect of noise uncertainty. Also we have simulated the performance of the CP-detector for the parameters of this section (i.e. $M = 40$) in Table \ref{table6}. Comparing Table \ref{table5} and Table \ref{table6}, we can see under all impairments excluding noise uncertainty, the energy detector has a better performance than the CP detector {compare column (iv) of Table \ref{table5} and column (i) of Table \ref{table6}}. When noise uncertainty is present, the performance of the energy detector degrades significantly compared to cyclic prefix detector and thus, in a snapshot setup, the CP-detector is more robust to these impairments (last columns of Tables \ref{table5} and \ref{table6}) than the energy detector.

\begin{table}
\begin{center}
\begin{tabular}[t]{|c|c|c|c|c|c|}\hline
$p_{fa}$ & \small{IQ Imbalance} & \small{IQ Imbalance} & \small{Noise Uncertainty} & \small{Timing, Frequency} & \small{All Impairments}
  \\
   & \small{(No Compensation)} & \small{(Compensation)} & & \small{ and IQ Imbalance} & \small{with compensation}\\
  &  &  & & \small{with compensation for IQ}& \small{for IQ}\\\hline
  $0.05$ & $0.9992$  &$0.9996$
  &$0.2785$ & $0.9995$ & $0.2563$
  \\\hline

  $0.025$ & $0.9981$  &$0.9989$
  &$0.1814$ & $0.9985$ & $0.1675$
  \\\hline

  $0.01$ & $0.9941$
  &$0.9968$
  &$0.1007$ & $0.9965$ & $0.0921$
   \\\hline
\end{tabular}\caption{Snapshot Energy Detector: $p_d$ under IQ-Imbalance, noise uncertainty and all impairments for the parameters of Sec. IV}
\label{table5}
\end{center}
\end{table}

\begin{table}
\begin{center}
\begin{tabular}[t]{|c|c|c|c|}\hline
  $p_{fa}$ & \small{CP-no Impairments} & \small{CP-timing, freq offsets}  & \small{All Impairments}
  \\
   & \small{(No Compensation)} & \small{and IQ Imbalance with} & \small{including noise uncertainty}\\
  & & \small{compensation for all of Sec. III} & \small{with compensation}\\
  & & & \small{for all of Sec. III}\\\hline
  $0.05$ & $0.9606$  
  &$0.7097$ & $0.5701$ 
  \\\hline

  $0.025$ & $0.9606$ 
  &$0.6173$ & $0.4680$ 
  \\\hline

  $0.01$ & $0.8728$
  &$0.4905$ & $0.3334$ 
   \\\hline
\end{tabular}\caption{Snapshot CP detector: $p_d$ under the parameters of Sec. IV}\label{table6}
\end{center}
\end{table}

\section{Cooperative Sequential Sensing of OFDM}
The advantages of spectrum sensing by cooperative means, i.e., using multiple nodes to sense the spectrum, are well known \cite{ref4}, \cite{ref9}. Furthermore, sequential detection is also known to perform better than snapshot detection. In this section, we apply cooperative sequential detection algorithms developed in \cite{ref10}, \cite{ref13}, \cite{ref19} for sensing of the OFDM signal in the setup of Section II. Interested readers are referred to \cite{ref10}, \cite{ref13}, \cite{ref19} for a more detailed introduction to sequential detection and its advantages.

We compare the performance of cooperative algorithms with different levels of impairments. DualCUSUM uses the well known CUSUM algorithm \cite{ref20} at the cognitive receivers as well as at the fusion node for detection of change (ON $\rightarrow$ OFF and OFF $\rightarrow$ ON of the primary). CUSUM is known to be optimal in different scenarios and uses the log likelihood ratio. Consequently, DualCUSUM has also been shown to perform very well (\cite{ref13}, \cite{ref19}). In the following, we use DualCUSUM in our present scenario and treat both energy detector and cyclic-prefix based detector simultaneously. We use the estimation schemes (wherever applicable) discussed in Section III and Section IV (suitably modified), overcoming the effects of different impairments.

In Tables \ref{table7} and \ref{table8} we provide the performance of DualCUSUM and its variants. The parameters used for simulations are described in Section V.E which also compares the algorithms in different scenarios.

\subsection{Dual CUSUM with No Impairments}
This is the ideal scenario where there are none of the impairments mentioned in Section III. For the cyclic prefix detector, correlation is done only over the length of samples corresponding to the cyclic prefix. Since all the parameters, including noise variance and received primary power are known, one can apply the DualCUSUM \cite{ref19} as explained briefly below. 

\begin{enumerate}
\item Each node $l$ computes the log likelihood ratio (LLR) $\xi_{j,l}$ of $R_r(j,l)$ in each slot $j(\geq 1)$ of $L_s$ samples as
\beq
R_r(j,l) = \real\left\{\frac{1}{L_c}\sum_{i=1}^{L_c}{X((j-1)L_s+i,l)X^*((j-1)L_s+L_d+i,l)} \right\},
\eeq
\beq
\label{eq_cp_dual_sigmal}
\xi_{j,l} = \frac{L_c(2 \sigma_{s,l}^2 R_r(j,l)-\sigma_{s,l}^4)}{2 \sigma_w^4}
\eeq
and computes the cumulative summation (CUSUM) 
\beq
W_{j,l}= (W_{j-1,l}+\xi_{j,l})^+, \quad W_{0,l}=0.
\label{cusum_recurse}
\eeq 
\item If the CUSUM crosses a threshold $\gamma$, it transmits a message $Y_{j,l} = b 1_{\{W_{j,i}>\gamma\}}$ to the fusion node (i.e., it sends a '$1$' with amplitude $b$). 
\item The fusion center receives $Y_j$ in slot $j$ where 
\beq
\label{phy_fusion}
Y_{j} = \sum_{l}{Y_{j,l}}+Z_j.
\eeq
and $\{Z_j\}$ is i.i.d. receiver noise with distribution $N(0,\sigma_M^2)$. 
\item The fusion node also runs CUSUM based on its input $Y_j$ by using the log likelihood ratio $\eta_j$ as follows:
\beq
F_k = (F_{k-1}+\eta_j)^+, \quad F_{0}=0, \quad \eta_j = \frac{2 Y_j bI-(bI)^2}{2 \sigma_M^2},
\eeq
where $I$ is a design parameter.
\item Fusion node finally declares change at time $\tau$ if $F_k$ exceeds a threshold $\beta$, i.e., 
\beq
\tau = \inf\{k:F_k > \beta\}. 
\eeq
\end{enumerate}

The parameters $\gamma$, $\beta$, $b$, $I$ affect the performance of the algorithm and the techniques developed in \cite{ref19} can be used to optimize performance. One computes $EDD= E[(\tau - T)^+]$ subject to the probability of false alarm $P_{FA} \leq \alpha \triangleq P[\tau < T]$. 

For the energy detector, the algorithm is the same as the above with minor modifications. The energy is computed as 
\beq
\label{energy_dualcusum}
V(j,l) =  \frac{\displaystyle\sum_{i=1}^{L_{s}} \arrowvert X((j-1)L_{s} + i,l)\arrowvert^{2}}{ML_{s}}
\eeq
and $\xi_{j,l}$ is the LLR computed with pre and post change distributions being $N(\sigma_{w}^{2},\sigma_{w}^{4}/ML_{s})$  and $N(\sigma_{s}^{2} + \sigma_{w}^{2},(\sigma_{s}^{2} + \sigma_{w}^{2})^{2}/ML_{s})$ respectively,
\beq
\label{eq_energy_dual_sigmal}
\xi_{j,l}=\frac{1}{2}\log\left(\frac{\sigma_{w}^{4}}{(\sigma_{s}^{2} + \sigma_{w}^{2})^{2}}\right) + \frac{\left(V(j,l)-\sigma_{w}^{2}\right)}{\sigma_{w}^{4}/ML_{s}} - \frac{\left(V(j,l)-(\sigma_{s}^{2} + \sigma_{w}^{2})\right)}{(\sigma_{s}^{2} + \sigma_{w}^{2})^{2}/ML_{s}}.
\eeq

For frequency selective fading, $V(j,l)$ in \eqref{energy_dualcusum} will not be i.i.d. pre and post change but will have some dependencies due to ISI (intersymbol interference). However this dependence will be weak because only a few symbols at the OFDM symbol boundary will get affected by the symbols of the previous OFDM symbol. Thus, one can continue to assume that $\{V(j,l\},j\geq1$ is an i.i.d. sequence, which is required to obtain the simplified algorithm described above. However the i.i.d. may not hold for the CP detector because CP resides near the boundary only. Thus, this case will require further consideration. However, we will see later, that in the sequential setup, energy detector significantly outperforms the CP detector in all possible scenarios we consider.

The performance of DualCUSUM has been obtained theoretically in \cite{ref19} and \cite{ref21}. It is a very efficient algorithm because it uses CUSUM at the local cognitive detectors and at the fusion node. Also, the local nodes transmit to the fusion node only if they are convinced that there is a change. This minimizes cognitive transmissions to the fusion node resulting in low transmit power consumption from cognitive nodes and low interference to the primary. Physical layer fusion (see \eqref{phy_fusion}) at fusion node (i.e., simultaneous transmissions from all cognitive users) further reduces this interference and also reduces the Expected Detection Delay ($EDD$). Its comparison with several other existing spectrum sensing algorithms is available in \cite{ref13}. 

\subsection{DualCUSUM with Timing Offset}
With an unknown timing offset, the decision statistic used at each node for the CP detector is as follows. First, the timing offset estimator of \eqref{eq7} is not preferred here, as under low SNR conditions, to minimize the estimation error, we need a large number $M$ of OFDM symbols \cite{ref17}. This will mean that the amount of memory required will be large. Thus, we propose the following. Each node runs $L_{d}$ CUSUMs for each possible timing offset of the primary. In slot  $j$, each node $l$ computes for $m \in \{0,1,2, ... , L_{d}-1\}$,
\begin{align}
&R_{r}(j,l,m) = Re \left( \dfrac{1}{L_{c}}\displaystyle\sum_{i=1}^{L_{c}}X((j-1)L_{s}+i+m,l)X^{*}((j-1)L_{s}+i+m+L_{d},l)\right), \nonumber \\
&\xi_{j,l,m} = \dfrac{(2\sigma_{s,l}^{2}R_{r}(j,l,m)-\sigma_{s,l}^{4})}{2\sigma_{w}^{4}/L_{c}},\quad W_{j,l,m} = (W_{j-1,l,m} + \xi_{j,l,m})^{+},  \\
&W_{j,l} = \max_{\{m\in 0,1,..L_{d}-1\}} W_{j,l,m}, \quad Y_{j,l} = b\mathbf{1}_{(W_{j,l} > \gamma)}.
\label{eq16}
\end{align}

This algorithm can be intuitively understood as follows. Before change, all the CUSUMs will typically be zero as $E[R_{r}(j,l,m)]=0$ before change. Once the primary arrives, the CUSUM corresponding to the correct timing offset $m=\theta$, will start increasing the fastest as it will capture the correct window of length $L_{c}$. This is similar to the Generalized Likelihood Ratio algorithm discussed in Section V.C for the unknown timing offset $\theta$, where $\theta \in \{0,1,\ldots,L_{d}-1\}$. More comments will follow in Section V.C.

None of the impairments at the secondary nodes studied above has any effect at the statistics of observations at the fusion node. We assume that the cognitive network knows its channel gains and has a better control over its system (this is a commonly made assumption in CR). Thus, the DualCUSUM at the fusion node remains unchanged. Furthermore, in our implementation, in slot 1, each node initially captures $L_{s} + L_{d}$ samples. From then onwards, each node captures only $L_{s}$ samples and uses the last $L_{d}$ samples from slot $j-1$ to calculate $R_{r}(j,l,m)$. It can be shown that $R_{r}(j-1,l,m)$  and $R_{r}(j,l,m)$ remain uncorrelated. This is because a sample in a set of consecutive $L_{d}$ samples will be correlated with some sample in slot $j-1$ or slot $j$, but not both. The performance of this algorithm is provided in Table \ref{table7}.

For the energy detector, since timing offset does not affect the decision statistics as discussed in Section IV.A, the algorithm remains the same as in Section V.A. Its performance is illustrated in Table \ref{table8}. A minor degradation in performance is observed. This is because in a change-detection setup, the presence of a timing offset implies that in the slot the primary comes on, the mean energy is less than $\sigma_w^2 + \sigma_s^2$. 

\subsection{GLR-CUSUM with Timing Offset, Frequency Offset and Primary Power Unknown}
Now we assume that $\sigma_{s,l}^{2}$ is unknown. Additionally, timing and   frequency offset could also be present. Thus, for the CP detector, we cannot use $R_{r}$ and need to use $R$ as the decision statistic instead (recall that $R_{r} = \real\{R\}$ and $R_{i} = \imag\{R\}$). It is easy to see that when frequency offset is present, post change, $R_{r}\sim N(\sigma_{s,l}^{2}\cos(2\pi \phi), \sigma_{w}^{4}/2L_{c})$ and $R_{i}\sim N(\sigma_{s,l}^{2}\sin(2\pi \phi), \sigma_{w}^{4}/2L_{c})$. Also, as we have no knowledge of primary signal power $\sigma_{s,l}^{2}$ , we now have a composite post change hypothesis, hence we use the Generalized Likelihood Ratio (GLR)-CUSUM algorithm \cite{ref10}. 

The GLR algorithm is briefly described as follows. Let $f_0$ be the density of the decision statistic $X_{j,l}$ before change and let $f_{\theta}$ be the density post change. Here $\theta$ is a parameter that characterizes the post-change distribution. In the case of CUSUM algorithm, the parameter $\theta$ is known and the CUSUM algorithm in slot $j$ can be described as
\beq
W_{j,l}=\max_{1 \leq s \leq k}\left(\sum_{i=s}^{k}\log\left(\frac{f_{\theta}(X_{i,l})}{f_{0}(X_{i,l})}\right)\right).
\label{cusum_eq}
\eeq
Equation (\ref{cusum_eq}) can be shown equivalent to (\ref{cusum_recurse}). In the case of GLR algorithm, $\theta$ is unknown, but $\theta \in \Theta \subseteq \Re$, where $\Re$ denotes the real line. Thus \eqref{cusum_eq} is changed to
\beq
W_{j,l}=\max_{1 \leq s \leq k}\left(\sup_{\theta \in \Theta}\sum_{i=s}^{k}\log\left(\frac{f_{\theta}(X_{i,l})}{f_{0}(X_{i,l})}\right)\right).
\label{glr_eq}
\eeq

In Section V.B, for the unknown timing offset scenario, the algorithm implemented can be described as
\beq
\tau_{\gamma,l}=\inf\{k:\max_{\theta \in \Theta}\max_{1 \leq s \leq k}\left(\sum_{i=s}^{k}\log\left(\frac{f_{\theta}(X_{i,l})}{f_{0}(X_{i,l})}\right)\right)>\gamma\}.
\label{parallel_cusum_eq}
\eeq
\noindent It should be noted that here $\sup$ is replaced with $\max$ as the set is finite and the $\max$ over the unknown parameter $\theta$ is moved outside. This is done because in the unknown timing offset scenario, keeping the $\max$ operation inside complicates the computations and requires much larger window sizes for the CP detector. This interchange possibly compromises the performance. However, from our simulations we will see that the degradation is negligible. 

Now, returning to the current impairments in OFDM, namely unknown frequency offset and primary signal power, the supremum is explicit. This is obtained by differentiating the likelihood ratio with respect to the unknown $\sigma_{s,l}^2, \phi$ and equating it to zero, and finally substituting the $\sigma_{s,l}^2, \phi$ which maximizes the likelihood ratio. Thus, the GLR test in combination with an unknown timing offset is given by
\begin{align}\label{eq:17}
&W_{j,l,m} = \max_{1\leq t \leq j} \dfrac{\left(\displaystyle\sum_{p=t}^{j}R_{r}(p,l,m)\right)^2 + \left(\displaystyle\sum_{p=t}^{j}R_{i}(p,l,m)\right)^2}{(j-t+1)\sigma_{w}^{4}/L_{c}}, \nonumber  \\
&\text{where $R_r$ and $R_i$ are the real and imaginary parts of} \nonumber \\
&R(j,l,m) =  \dfrac{1}{L_{c}}\displaystyle\sum_{i=1}^{L_{c}}X(jL_{s}+i+m,l)X^{*}(jL_{s}+i+m+L_{d},l), \quad m \in \{0,1,2, ... , L_{d}-1\}.
\end{align}
The above equation can be intuitively understood as follows. Before the change mean of both $R_r$ and $R_i$ are zero, and thus $W_{j,l,m}$ will be close to zero. After the change, for the $m=\theta$ (i.e., for the CUSUM corresponding to the correct timing offset) since the mean is nonzero, $W_{j,l,m}$ will keep increasing with $j$, thus eventually detecting the change. The rest of the steps at each secondary node are the same as in Section V.A. At the fusion node, the DualCUSUM operation remains unchanged. The computations in the GLR algorithm can be limited to a finite window as suggested in \cite{ref10}.

For the energy detector, the frequency offset does not affect the performance but due to lack of knowledge in primary power we need to use the GLR algorithm.  The energy in each slot is 
\beq
V(j,l) = \displaystyle\sum_{i=1}^{L_{s}} \frac{ \arrowvert X((j-1)L_{s} + i,l )\arrowvert^{2} - \sigma_{w}^{2}}{ ML_{s}}.
\eeq
(Here subtraction by $\sigma_{w}^{2}$ is performed for convenience, for making mean zero before change). At slot, $j$, the GLR algorithm is as follows:
\begin{align}\label{eq:18}
&W_{j,l} = \max_{1\leq i \leq j} A_{i,j,l}, \quad \text{where} \nonumber \\ 
&A_{i,j,l} = \dfrac{\displaystyle\sum_{p=i}^{j}V(p,l)^{2}}{2\sigma_{w}^{4}/ML_{s}} - \dfrac{\displaystyle\sum_{p=i}^{j}(V(p,l)-
 \theta_{1}(i,j,l))^2}  {2(\theta_{1}(i,j,l)+\sigma_{w}^{2})^{2}/ML_{s}} + \dfrac{1}{2}\log\left(\dfrac{\sigma_{w}^{4}}{(\theta_{1}(i,j,l) + \sigma_{w}^{2})^{2}}\right).
\end{align}

And $\theta_{1}(i,j,l)$ is obtained by solving the quadratic equation for $\theta_{1}$
\begin{align}\label{eq:19}
& (j-i+1)\theta_{1}^{2} + \theta_{1}\left(2(j-i+1)\sigma_{w}^{2} + ML_{s}(j-i+1)\sigma_{w}^{2} + S_{i,j,l}\right) \nonumber \\ &~~~~~~~~~~~-\left(ML_{s}SQ_{i,j,l}+ML_{s}\sigma_{w}^{2}S_{i,j,l} - (j-i+1)\sigma_{w}^{4}\right)= 0.
\end{align}
\noindent where $S_{i,j,l}=\displaystyle\sum_{p=i}^{j}V(j,l)$ and $SQ_{i,j,l} = \displaystyle\sum_{p=i}^{j}V(j,l)^{2}$. In the above equation $\theta_{1}(i,j,l)$ denotes an estimate of $\sigma_{s,l}^2$ (assuming primary has come ON in slot $i$) and is chosen from $\theta_1 \in [0,\infty)$. The quadratic equation for $\theta_{1}$ was obtained by simply differentiating the likelihood ratio w.r.t $\theta_{1}$ and setting equal to zero. The rest of the steps at a secondary node are same as in DualCUSUM, and fusion node continues to use the CUSUM algorithm. The performance of this algorithm is illustrated in Table \ref{table8}.  

\subsection{Algorithms for all Impairments}
We assume that all the above mentioned impairments (including IQ imbalance) could be present and $\sigma_{w}^{2}$ and $\sigma_{s}^{2}$  are unknown to the secondary nodes. For the CP detector, while we can extend the GLR test to cover this scenario as well, we have found via simulations, that it is better to first compensate for the IQ-imbalance in each slot using (\ref{eq12}) and \eqref{eq13}. Then we estimate the noise power as 
\beq
\widehat\sigma_{w,j,l}^{2} = \frac{\displaystyle\sum_{p=1}^{j}\sum_{k=1}^{L_{s}}\arrowvert X((p-1)L_{s} + k, l) \arrowvert^{2}}{jL_{s}}.
\eeq
This approximation is valid under low SNR assumptions, as we assume the same value for the variance under either hypothesis. Now, since the IQ imbalance can be assumed to have been corrected and we have an estimate of noise power $\widehat\sigma_{w,j,l}^{2}$, we can use the setup of Section V.C for the other impairments (timing offset, frequency offset and received primary power). Thus, for CP detector each node, does the same as in Section V.C using the estimated noise power $\widehat\sigma_{w,j,l}^{2}$ in slot $j$.

For the energy detector, the uncertainty in noise power requires the modified GLR (MGLR) algorithm \cite{ref10}. In a CP based detector this was not required as it performed detection of change in the mean of a Gaussian signal, and before the change, the mean was known to be zero. Thus, the unknowns are post-change mean, and the variances before and after change. Since, under low SNR, variance approximately remains same before and after the change, GLR can be used as discussed in the previous paragraph. But in case of the energy detector, while the unknown variance, is approximately the same (under low SNR) before and after change, the mean both before and after change is also unknown and thus we need to use a modified version of GLR (MGLR) algorithm. To clarify a bit more, in comparison to (\ref{glr_eq}) the MGLR equation will look as 
\begin{align} \label{mglr_eq}
\tau_{\gamma,l}=inf\{k:\max_{1 \leq s \le k, s \geq M^{*}}\left(\sup_{\theta^{'} \in \Theta}\sum_{i=1}^{s}\log\left(f_{\theta^{'}}(X_{i,l})\right)\right)+\left(\sup_{\theta^{''} \in \Theta}\sum_{i=s+1}^{k}\log\left(f_{\theta^{''}}(X_{i,l})\right)\right) \nonumber \\
-\left(\sup_{\theta \in \Theta}\sum_{i=1}^{k}\log\left(f_{\theta}(X_{i,l})\right)\right)>\gamma\}
\end{align}

\noindent where $\theta^{'}$ and $\theta^{''}$ are possible parameters before and after change, and $\theta$ evaluates the possibility that there is no change. The MGLR approach was first outlined in \cite{ref10}. The method relies the presence of $M^{*}$ samples pre-change. Loosely speaking, the initial set of samples where $H_{0}$ is the true hypothesis, helps in estimating the unknown parameters for subsequent sequential detection of change in the presence of impairments. The value of $M^{*}$ depends upon the minimum SNR at which we need to detect reliably and the $P_{FA}$ desired. The MGLR algorithm for the energy detector becomes 
\begin{align}\label{eq_full_mglr}
&W_{j,l} = \max_{M^{*} \leq i < j}A_{i,j,l} 1_{\theta_1(i+1,j,l)>\theta_1(1,i,l)},\quad \text{where}, \nonumber \\
&V(j,l) = \dfrac{\displaystyle\sum_{i=1}^{L_{s}}\arrowvert X(jL_{s}+i,l)\arrowvert^{2}}{ML_{s}}, \nonumber \\
&A_{i,j,l} = B_{1}^{i}(l) + B_{i+1}^{j}(l) - B_{1}^{j}(l), \quad B_{a}^{b}(l) = \dfrac{\displaystyle\sum_{p=a}^{b}\left(V(p,l)-\theta_{1}(a,b,l)\right)^{2}}{2\theta_{1}(a,b,l)^{2}/ML_{s}}.
\end{align}
And $\theta_{1}(a,b,l)$ is obtained by solving the quadratic equation for $\theta_{1}$
\begin{align}
\label{eq_full_mglr_2}
&(b-a+1)\theta_{1}^{2} + \theta_{1}ML_{s}S_{a,b,l}-ML_{s}SQ_{a,b,l}=0.
\end{align}
\noindent In the equation \eqref{eq_full_mglr_2} $\theta_1(1,i,l)$ is an estimate of $\sigma_{w,l}^2$ and $\theta_1(i+1,j,l)$ is an estimate of $\sigma_{s,l}^2 + \sigma_{w,l}^2$, assuming primary has come on at slot $i+1$. $\theta_1(1,j,l)$ is an estimate of $\sigma_{w}^2$ assuming primary has not come on. The rest of the steps at a secondary node are the same as in DualCUSUM and fusion node continues to use CUSUM. The performance of this algorithm is illustrated in Table \ref{table8}. 

The condition in \eqref{eq_full_mglr} is for detecting OFF$\rightarrow$ON, i.e., we are detecting an increase in signal power. The condition needs to be reversed for detecting ON$\rightarrow$OFF \cite{ref10}.

\subsection{Performance Comparison}
In this subsection, we compare the performance of the above algorithms. There are 5 nodes. The SNR at each node is $-10$dB. Wherever applicable, $\phi=0.1, \Delta\phi = 10^{o},\epsilon=0.2, \theta=10$. The change time $T$ (in units of OFDM symbols) is assumed to have a geometric distribution with parameter $\rho = 0.004$. For different values of $P_{FA}$, $EDD$ in units of OFDM symbols is shown in Table \ref{table7} for CP-based detectors and in Table \ref{table8} for energy-detector based algorithms. 

For comparison, we have also simulated a cooperative snapshot detector for both CP and energy detectors. CP detector captures $M=50$ OFDM symbols of data and detects the signal in the presence of all impairments and compensating for the same using the steps in Section III.E. The energy detector captures $M=5$ OFDM symbols and detects the signals in presence of all impairments. Compensation is done for IQ imbalance. The values of $M$ chosen for the two snapshot detectors are chose to minimize $EDD$ in each case for a given $P_{FA}$.
Each node sends a $1$ or $0$ according to whether $H_{1}$ or $H_{0}$ is chosen. The fusion center uses the AND rule to decide between $H_{0}$ or $H_{1}$ as the AND rule works the best in the present setup. For the snapshot detector, we assume that the fusion node has no noise. 

\begin{table}
\centering
\begin{tabular}{|c|c|c|c|c|c|}
\hline
$P_{FA}$ & (IV.A) & (IV.B) & (IV.C) & (IV.D) & Snapshot \\
 &  &  & & \small(all impairments) & \small(all impairments) \\
\hline
0.1 & 10.15 & 18.27 & 24.71 & 28.15 & 64.16\\
\hline
0.075 & 11.43 & 19.82 & 28.07 & 31.01 & 67.46\\
\hline
0.05 & 12.6 & 22.09 & 31.42 & 34.95 & 72.35\\
\hline
\end{tabular}
\caption{CP based co-operative spectrum sensing algorithms.}
\label{table7}
\end{table}

\begin{table}
\centering
\begin{tabular}{|c|c|c|c|c|c|}
\hline
$P_{FA}$ & (IV.A) & (IV.B) & (IV.C) & (IV.D) & Snapshot\\
& & & & \small(all impairments) & \small(all impairments)\\
\hline
0.1 & 5.22 & 5.43 & 7.73 & 10.15 & 349.13\\
\hline
0.075 & 5.61 & 5.91 & 8.52 & 11.43 & 438.03\\
\hline
0.05 & 6.41 & 6.46 & 9.19 & 12.6 & 623.58\\
\hline
\end{tabular}
\caption{Energy detector based co-operative spectrum sensing algorithms.}
\label{table8}
\end{table}

We see that, as the amount of uncertainty increases, the performance degrades for both the detectors. Also, from last two columns, we see that the sequential setup provides significant performance gains over the snapshot detector (even though for the snapshot detector we have assumed no noise at the fusion node) for both the CP-detector and the Energy-detector. Also, comparing the energy detector and the CP-detector we can clearly see that in a change-detection setup, energy detector significantly outperforms the CP detector (by comparing the columns labeled IV.D in both the tables) under all scenarios. (In this example $M^{*}$ for the MGLR was chosen as 50 OFDM symbols). But the snapshot energy detector shows significant degradation under noise uncertainty. 

\section{Conclusions} \label{sec:Conc}
We have considered the problem of spectrum sensing of OFDM signals using cyclic prefix based and energy based detectors. We have analyzed the effect of some typical impairments like timing and frequency offset, IQ-imbalance and transmit/noise power uncertainty and presented techniques to modify the detectors to work under these impairments. We have also proposed cooperative sequential change detection based algorithms and overcome the effects of these impairments in that setup also. We have shown that sequential detection improves the performance significantly as against fixed sample size detectors. It is also shown that the sequential energy detector significantly outperforms the CP detector under all impairments but the snapshot energy detector performs worse than the CP detector under noise power uncertainties. 

Most of these detectors will work under time varying multipath frequency selective fading also. In future work we will verify these claims via simulation and also consider frequency selective fading under different impairments discussed in this paper.


\end{document}